# Fast-response silicon photonic microheater induced by parity-time symmetry breaking

Yanxian Wei,[1] Junwei Cheng,[1] Yilun Wang,[1] Hailong Zhou,[1, 2, 3*] Jianji Dong,[1, #] Dongmei Huang,[3, 4] Feng Li,[2, 3] Ming Li,[6] P. K. A. Wai,[2, 3, 5] and Xinliang Zhang[1]

[1]Wuhan National Laboratory for Optoelectronics, Huazhong University of Science and Technology
[2]Photonics Research Centre, Department of Electronic and Information Engineering, The Hong Kong Polytechnic University
[3]The Hong Kong Polytechnic University Shenzhen Research Institute, Shenzhen 518057, China
[4]Photonics Research Centre, Department of Electrical Engineering, The Hong Kong Polytechnic University, Hong Kong
[5]Department of Physics, Hong Kong Baptist University, Kowloon Tong, Hong Kong
[6]State Key Laboratory on Integrated Optoelectronics, Institute of Semiconductors, Chinese Academy of Sciences, Beijing 100083, China
Corresponding author:[*] hailongzhou@hust.edu.cn; [#]jjdong@mail.hust.edu.cn

**Abstract:** Thermo-optic microheater is indispensable in silicon photonic devices for smart and reconfigurable photonic networks. Much efforts have been made to improve the metallic microheater performance in the past decades. However, because of the metallic nature of light absorption, placing the metallic microheater very close to the waveguide for fast response is impractical and has not been done experimentally. Here, we experimentally demonstrate a metallic microheater placed very close to the waveguide based on parity-time (PT) symmetry breaking. The intrinsic high loss of metallic heater ensures the system will operate in the PT-symmetry-broken region, which guarantee the low loss of light in the silicon waveguide. Moreover, heating at a close range significantly reduces the response time. A fast response time of ~1 μs is achieved without introducing extra loss. The insertion loss is only 0.1 dB for the 100-μm long heater. The modulation bandwidth is 280 kHz, which is an order of magnitude improvement when compared with that of the mainstream thermo-optic phase shifters. To verify the capability of large-scale integration, a 1×8 phased array for beam steering is also demonstrated experimentally with the PT-symmetry-broken metallic heaters. Our work provides a novel design concept for low-loss fast-response optical switches with dissipative materials and offers a new approach to enhance the performance of thermo-optic phase shifters.

## Introduction

Information explosion has become the major challenge to communication systems in recent years. As a promising solution to handle the tremendous amount of data, integrated silicon photonics have the advantages of high speed, compactness and low loss. Modulators are indispensable in constructing an on-chip communication system. There are many on-chip modulation methods to implement optical signal processing such as electro-optic modulations[1, 2, 3], thermo-optic modulations[4, 5, 6, 7, 8] and electro-absorption modulations[9]. Especially, thermo-optic modulations with advantages of easy fabrication, low cost and compactness have been applied in many optical devices [10, 11, 12, 13]. But the bandwidth is limited by the conduction and dissipation speed of heat. By applying methods such as doping technique[14, 15, 16], slow light effect[17] or two-dimensional materials[18, 19, 20], the performance of microheaters could be greatly improved. But for the widely used metallic heaters, the performance is still limited owing to strong optical absorption. Much researches have been focused to improve the conduction and dissipation speed of metallic heaters, such as digging an air trench under the thermo-optic waveguide[21], using laterally supported suspended arms[6], or just optimizing the size and position of commercial thermo-optic heater[5]. However, the conduction and dissipation speed is still seriously limited, because the metallic heaters are placed far away from the waveguide to avoid light absorption. Meanwhile, the thick oxide layer or air layer between the heater and waveguide significantly weakens the conduction and dissipation speed[5].

Traditionally, absorption loss of the metallic heater is regarded as a disadvantage. Recently, an alternate view on dissipation emerges as dissipation can be a powerful manipulation tool, even enables some counterintuitive characteristics as a result of parity–time (PT) symmetry breaking[22, 23, 24]. Non-Hermitian systems usually exhibit PT symmetry, i.e., the system is invariant under the parity reversal and time reversal. A PT-symmetric photonic system usually contains two coupled physical parameters with balanced gain and loss. PT symmetry can be broken and restored by manipulating the gain and loss of the system. These concepts had been widely studied to enhance the performance of the optical devices[25, 26], such as optoelectronic oscillators[27, 28], single-mode microring lasers[23, 29, 30, 31], sensing[32], unidirectional invisibility[33, 34] and asymmetrical modes transmission[35, 36, 37, 38, 39]. Specially, the PT symmetry can be realized in loss-only structures, as passive PT-symmetric systems. The ability to enhance the gain coefficient of one guided mode was first highlighted in a passive PT-symmetric experiment by introducing additional loss to another waveguide [24]. The loss-induced gain idea was subsequently extended to microwave and optical resonators[23]. Inspired by the properties of passive PT-symmetric systems, it is possible to apply the principle of PT symmetry to guide the design of thermo-optic phase shifter with the metal placed very close to the waveguide, to enhance the performance but without introducing extra loss.

In this paper, we designed and demonstrated a low-loss silicon phase shifter with fast-response (~1 μs) metallic heater based on the principle of PT symmetry breaking. We then experimentally verified the performance of thermo-optic phase shifter using a Mach-Zehnder interferometer (MZI) structure. Both lateral heating scheme (the metal is laterally placed) and top heating scheme (the metal is top placed) are demonstrated. A bandwidth up to 280 kHz is obtained. The rise time and decay time are 1.35 and 1.15 μs respectively for the lateral heating scheme and are 2.05 and 1.35 μs for the top heating scheme. The measured power for $\pi$ phase shift ranges from 17 to 22 mW for a 100-μm long heater. Compared to the commercial metallic heaters, the modulation speed is improved by more than ten folds and the loss induced by the metallic heater is negligible. We also successfully demonstrate a 1×8 optical phased array for beam steering based on the proposed microheaters, verifying the potential for large-scale integration. The proposed microheaters have great potential for the fast-response phase shifter applications such as optical switch, thermo-optic modulation and optical filtering. Our work also further expand the perspective of light control based on dissipative materials, paving a way for practical applications of PT symmetry.

## Results
### Theory of PT-symmetric systems
For a two-waveguide-coupled PT-symmetric system, the mode evolution with an increasing loss of the waveguide 2 (WG2) is shown in Figs. 1(a) to 1(c). We use the coupled-mode theory to analyze this system and the eigenvalues are given by $\varepsilon = \pm\sqrt{\Omega^2 - \gamma^2/4}$, where $\Omega$ is the coupling coefficient and $\gamma$ is the gain/loss difference of two waveguides. Here, $|\Omega|>|\gamma|/2$ indicates that the PT symmetry is not broken (Fig. 1(a)), while $|\Omega|<|\gamma|/2$ indicates the PT symmetry is broken (Fig. 1(b)). The coupled waveguides system can be switched between the PT symmetry unbroken state and broken state by adjusting the loss parameter $\gamma$. When the loss of the WG2 is sufficiently large and deep into the PT-symmetry-broken range as shown in Fig. 1(c), the light will be confined in WG1 with negligible loss.

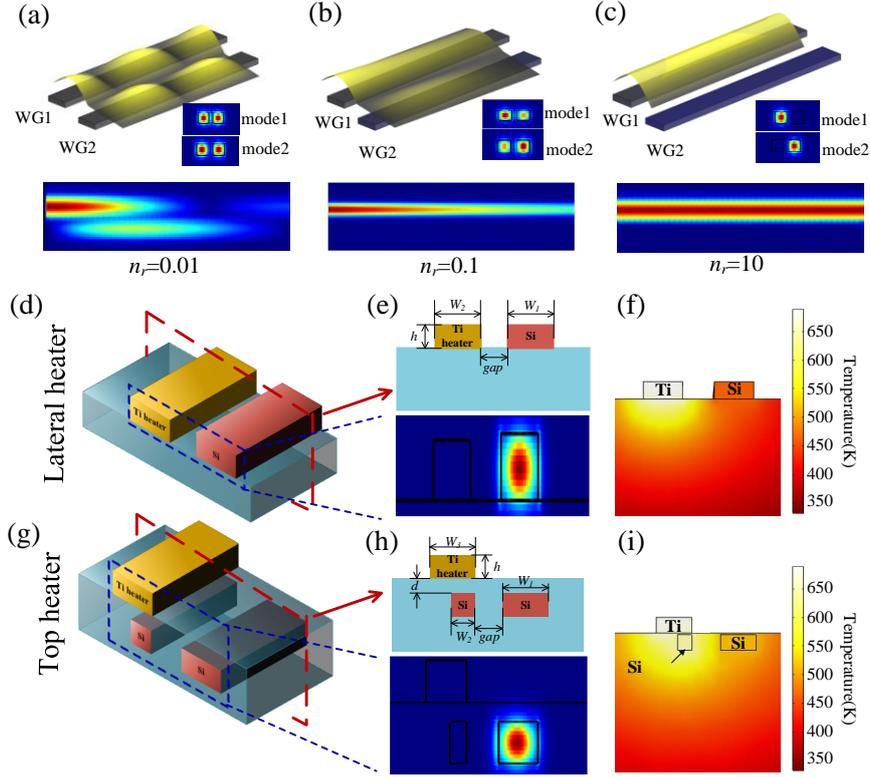

Fig.1. Schematic diagrams of two-waveguide-coupled PT-symmetric systems and the design of the microheaters. (a-c) Propagating fields and eigenmode distributions and for (a) unbroken, (b) broken and (c) deeply-broken PT symmetry. (d-f) The lateral heating schematic. $W_1=W_2=500$ nm, $h=220$ nm, $gap=400$ nm. (g-i) The top heating schematic. $W_1=W_3=500$ nm, $h=220$ nm, $W_2=gap=220$ nm, $d=200$ nm.

Intuitively, the waveguide should be high-loss if the dissipative material, such as metal, is placed very close to the waveguide, but such is not the case in a broken PT-symmetric system. We consider a passive system in which WG1 is lossless and WG2 has a large loss, i.e., $\gamma_1=0, \gamma_2\ll 0$. In the simulations, the eigenvalues of the two-waveguide-coupled system are calculated by using the finite-difference-time-domain method. The waveguide loss is proportional to the imaginary part of the refractive index ($n_r$). We apply different $n_r$ and analyze the evolution of the eigenvalues. Theoretically, when $n_r$ increases ($n_r = 0.01$) but is still within the unbroken PT-symmetric region, the light field in the system will oscillate between the two waveguides with a high global loss (Fig. 1(a)). However, when $n_r$ increases further ($n_r = 0.1$) until the system is in the PT-symmetry-broken states, the light field in the system will mainly be confined in WG1 with a relatively low loss (Fig. 1(b)). Finally, when the loss of the WG2 ($n_r = 10$) is well beyond the symmetry breaking threshold, where the eigenvalues are expressed as $\varepsilon \approx \pm \frac{\gamma}{2} i$, the light propagates only in WG1 with a negligible loss (Fig. 1(c)). From Figs. 1(a) to 1(c), the light in two waveguides will be not coupled when the loss of the high-loss waveguide is sufficient high. Thus the PT symmetry breaking phenomenon can be used to design a loss-induced thermo-optic phase shifter with the metallic heater placed close to the waveguide. The imaginary part of the refractive index of the metal is chosen to be higher than the PT symmetry breaking threshold so that the phase shifter is working in the symmetry breaking phase.

**Design of loss-induced silicon microheaters**
We design the phase shifter with a titanium (Ti) heater, which can heat the silicon (Si) waveguide at a very close range. The complex refractive index of Ti is $n_{Ti} = 4.04 + 3.82i$. The imaginary part of $n_{Ti}$ is located in the PT-symmetry-broken range. From the aforementioned analysis, light propagating in a Ti-Si coupled system can be confined in the Si waveguide without any additional loss. Figs. 1(d) to 1(i) show two designs of the microheater. The first design consists of a Si waveguide, shown in Fig. 1(d). A Ti heater is placed laterally along a Si waveguide at a spacing of 400 nm. The Ti heater can be treated as a high-loss waveguide in the PT-symmetric system. Thus, the light will be confined in the Si waveguide with a low loss. Fig. 1(e) shows the parameters and the simulated gain eigenmode of the lateral heating scheme. As we can see, the light field is confined in the Si waveguide and the loss of transverse electrical (TE) mode is ~0.016 dB for a 100 μm heater. Figure 1(f) shows the thermal distribution. The second design consists of two waveguides. The Ti heater is located on top of the left Si waveguide, thus it is a top heating scheme. Similar to Fig. 1(e), Fig. 1(h) shows that the light field is confined in the Si waveguide, and the loss of TE mode is about 0.01 dB for a 100-μm long heater. The gap between the Ti heater and Si waveguide is 400 nm for both schemes, which is rather small when compared to typical commercial values of 2~3 microns. Fig. 1(i) shows the thermal distribution of the second design.

The heat conduction of the two designs are different. In the lateral heating scheme, the heat mainly conducts through the $SiO_2$ buried oxide (BOX) layer, while in the top heating scheme, the heat conducts through the top $SiO_2$. The Si waveguide temperatures are 475 K and 525 K for the lateral heating scheme and top heating scheme respectively when the applied power to the metallic heaters is 20 mW. As we can see, the efficiency of lateral heating scheme is slightly lower than that of the top one, it is likely due to thermo-conduction between heater and waveguide of lateral heating scheme is worse than that of top heating scheme, while both schemes have sufficient conduction efficiencies for the thermo-optic modulation.

**Experimental demonstration of silicon microheater**
Based on the design of loss-induced silicon microheaters, we fabricate an MZI optical switch which contains the microheater. The microscope images of fabricated lateral heating and top heating chips are shown in Figs. 2(a) and 2(d) respectively, and the scanning electron microscopes (SEMs) images of the microheaters are shown in the insets at the bottom. To determine the insertion loss induced by the heater, we measure the spectra of the MZIs with our microheaters and reference MZIs without heaters. The metallic absorption losses induced by the heater can be evaluated by comparing the spectra and are about 0.1 dB and 0.2 dB at a length of 100 μm for the two structures respectively. The difference between the experimental and simulated loss is caused by the measurement and fabrication errors. For example, some metal fragments are closer to the waveguide than design value, which will introduce an additional loss, as shown in the SEM images of Fig. 2(a).

In terms of the lateral heating scheme, Fig. 2(b) presents the transmission spectra when different driving power is applied to the heater in the lateral heating scheme. We can see that the transmission spectrum is shifted approximately linearly with the driving power and the power for $\pi$ phase shift is about 22 mW. Fig. 2(c) shows a tuning efficiency of 0.0324 nm/mW. Figs. 2(d) to 2(f) show the experimental characterization of the top heating structure similarly to that shown in Figs. 2(a) to 2(c). The power for $\pi$ phase shift is about 17 mW, which is slightly less than that of lateral heating scheme. The results agree with the prediction from simulations.

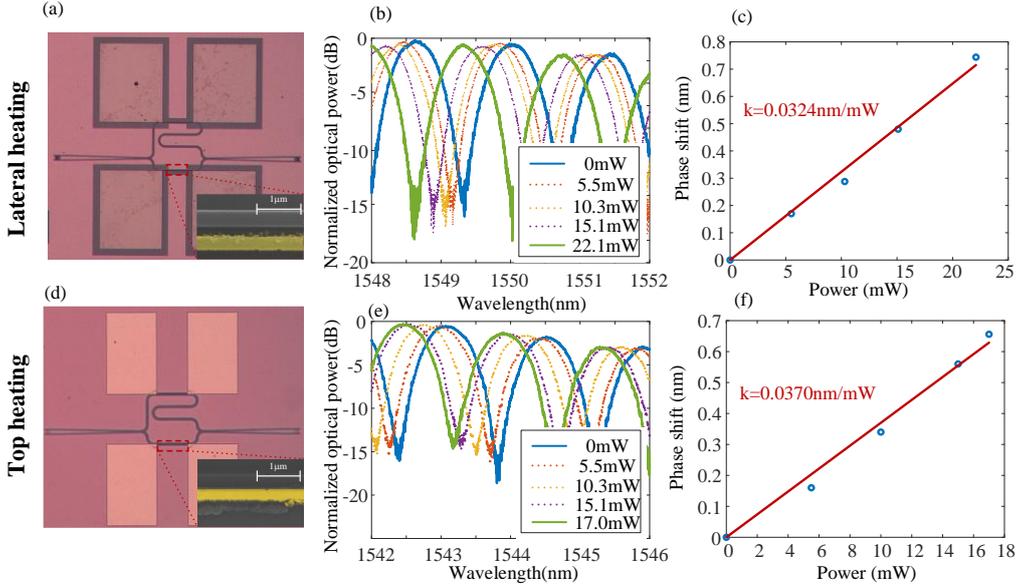

Fig. 2. Images of chips and experimental characterization of silicon microheaters. Microscope images of MZIs and the zoom in SEM images of microheaters for the (a) lateral and (d) top heating scheme. Transmission spectra for different driving power of the (b) lateral and (e) top heating scheme. Phase shift versus heating power for the (c) lateral and (f) top heating scheme.

We use a 10 kHz square-wave driving signal to characterize the response speed of heaters, the results are shown in Figs. 3(a) and 3(b) for the lateral and top heating scheme respectively. As shown in Fig. 3(a), the rise time and decay time are 1.35 μs and 1.15 μs for the lateral heating scheme, respectively, which correspond to a response speed of about 280 kHz. The performance of the top heating scheme is similar to that of the lateral heating scheme. From Fig. 3(b), the rise time and decay time are 2.05 μs and 1.35 μs, respectively, which correspond to a response speed of about 205 kHz. Fig. 3(c) gives the amplitude-frequency response curves for the two structures. We can see that the measured 3-dB bandwidths of the lateral and top heating schemes are 280 kHz and 205 kHz respectively, which agrees with the results obtained by using a 10 kHz square-wave driving signal. Both schemes have high response speeds which is an order of magnitude improvement when compared with the commercially available thermo-optic phase shifters.

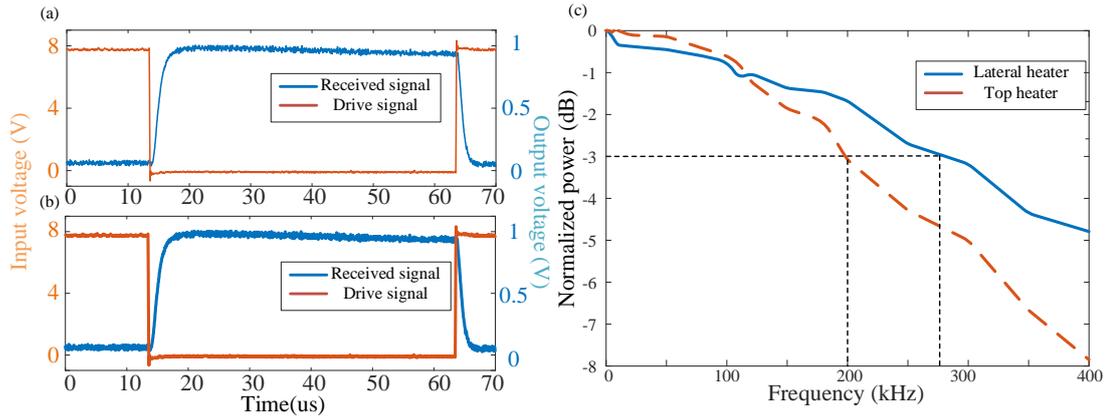

Fig.3. Experimental characterization of response speed of the devices. Response of 10 kHz square-wave for the (a) lateral and (b) top heating scheme. (c) Amplitude-frequency response curves of the two microheaters.

To demonstrate the potential of large-scale integration of the proposed microheaters, we designed and fabricated a one-dimensional optical phased array for beam steering. Fig. 4(a) shows the microscope image of the phased array. Light is coupled into the chip from the right side by a grating and then separated into eight branches by Y-branch waveguides. The phase of light in each branch can be adjusted by a top heating microheater and finally the light radiates into free space by a grating array. The output gratings have a width of 4 μm and a spacing of 0.5 μm . The chip is packaged as shown in Fig. 4(b). The input fiber is packaged with a 45° reflective crystal head and the chip is glued on PCB. The electrodes on the chip are connected to the PCB by wire bonding.

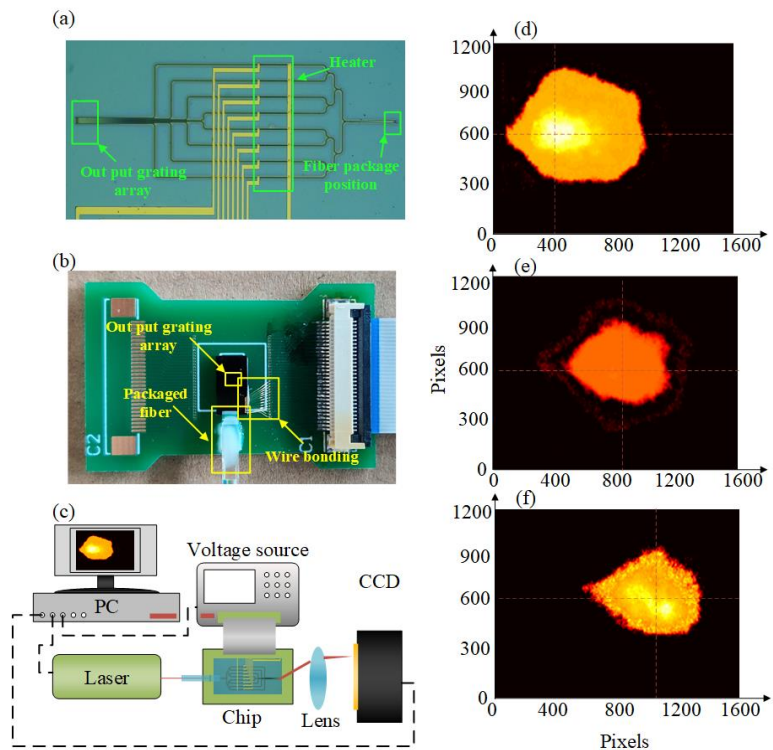

Fig.4. Optical phased array. (a) Microscope image of the chip. (b) Packaged chip. (c) Experimental setup. (d-f) Photos of the light spots acquired by the CCD for different voltages applied on the heaters.

Fig. 4(c) shows the experimental setup. A tunable laser is used to generate the 1550 nm input light and the chip is adjusted by a voltage source. The light output from the chip radiates into free space and a lens with 5 cm focal length is used to acquire the far field. A CCD camera is used to detect the light. All the instruments are controlled by a personal computer (PC). The light spots received by the CCD camera are shown in Figs. 4(d), 4(e) and 4(f). The spot location can be varied in the horizontal direction by adjusting the voltages applied on the heaters. A beam-steering range of ± 3.45° is realized. The performance of phased array can be enhanced by carefully designing the period and number of gratings.

## Conclusions

In conclusions, we theoretically analyze PT symmetry breaking in two-waveguide-coupled systems and apply the principle to design close-range microheaters. Owing to the narrow spacing between heater and waveguide, we observe a 3-dB bandwidth up to 280 kHz experimentally. The rise time and decay time are measured to be 1.35 μs and 1.15 μs respectively for the lateral heating scheme, and 2.05 μs and 1.35 μs respectively for the top heating scheme. To the best of our knowledge, it is the fastest modulation speed of metallic microheaters ever observed without doping process. Compared to other concepts of microheater[6, 17, 40, 41], the proposed microheaters have advantage on large-scale integration, which is demonstrated by a 1×8 phased array. The proposed loss-induced microheater provides a promising solution to improve the modulation speed of commercially available thermo-optic devices and have great potential in the future photonics integrated circuits.

## Acknowledgements

This work was partially supported by the National Key Research and Development Project of China (2018YFB2201901), the National Natural Science Foundation of China (61805090, 62075075), Research Grants Council of Hong Kong SAR (PolyU152241/18E).

## Authors' contributions

Y.X.W. and H.L.Z. conceived the idea. Y.X.W. performed the numerical simulations and designed the device. Y.X.W. and J.W.C. performed the experiments. Y.X.W. and Y.L.W. fabricated the chip. Y.X.W., F.L., D.M.H., P.K.A.W. and L.M. discussed the results. J.J.D and X.L.Z supervised the study. All authors contributed to the writing of the paper.

## Data availability

All the data supporting this study are available in the paper and Supplementary Information. Additional data related to this paper are available from the corresponding authors upon request.

## References

1. Cazier N, Sadeghi P, Chien M-H, Shawrav MM, Schmid S. Spectrally Broadband Electro-Optic Modulation with Nanoelectromechanical StringResonators. *Opt. Express* **28**, 12294-12301 (2020).

2. Haffner C, *et al.* Low loss Plasmon-assisted electro-optic modulator. *Nature* **556**, 483-486 (2018).

3. Jin M, Chen J-Y, Sua YM, Huang Y-P. High-extinction electro-optic modulation on

lithium niobate thin film. *Opt. Lett.* **44**, 1265 (2019).

4. SungWon, Chung, Makoto, Nakai, Hossein, Hashemi. Low-power thermo-optic silicon modulator for large-scale photonic integrated systems. *Opt. Express*, 13430-13459 (2019).

5. Maxime, *et al.* Optimization of thermo-optic phase-shifter design and mitigation of thermal crosstalk on the SOI platform. *Opt. Express*, 10456-10471 (2019).

6. Chen K, Duan F, Yu Y. Performance-enhanced silicon thermo-optic Mach–Zehnder switch using laterally supported suspended phase arms and efficient electrodes. *Opt. Lett.* **44**, 951-954 (2019).

7. Chang Y-C, Roberts SP, Stern B, Lipson M. Resonance-Free Light Recycling. Preprint at https://ui.adsabs.harvard.edu/abs/2017arXiv171002891C (2017).

8. Yuhan YAO ZC, Jianji DONG, Xinliang ZHANG. Performance of integrated optical switches based on 2D materials and beyond. *Front. Optoelectron.* **13**, 129-138 (2020).

9. Nevou L, *et al.* Short-wavelength intersubband electro-absorption modulation based on electron tunneling between GaN/AlN coupled quantum wells. *Appl. Phys. Lett.* **90**, 3722 (2007).

10. Hailong Z, *et al.* Silicon-based polarization analyzer by polarization-frequency mapping. *Apl Photonics* **3**, 106105-106115 (2018).

11. Zhou HL, Zhao YH, Wei YX, Li F, Dong JJ, Zhang XL. All-in-one silicon photonic polarization processor. *Nanophotonics* **8**, 2257-2267 (2019).

12. Zhou H, Zhao Y, Wang X, Gao D, Dong J, Zhang X. Self-Configuring and Reconfigurable Silicon Photonic Signal Processor. *ACS Photonics* **7**, 792-799 (2020).

13. Zhou H, *et al.* Chip-Scale Optical Matrix Computation for PageRank Algorithm. *J. Sel. Top. Quant.* **26**, 1-10 (2020).

14. Li X, Xu H, Xiao X, Li Z, Yu Y, Yu J. Fast and efficient silicon thermo-optic switching based on reverse breakdown of pn junction. *Opt. Lett.* **39**, 751-753 (2014).

15. Ding J, *et al.* Ultra-low-power carrier-depletion Mach-Zehnder silicon optical modulator. *Opt. Express* **20**, 7081-7087 (2012).

16. Asheghi M, Kurabayashi K, Kasnavi R, Goodson KE. Thermal conduction in doped single-crystal silicon films. *Journal of Applied Physics* **91**, 5079-5088 (2002).

17. Yan S, *et al.* Slow-light-enhanced energy efficiency for graphene microheaters on silicon photonic crystal waveguides. *Nat. Commun.* **8**, 14411 (2017).

18. Wei KK, *et al.* All-optical PtSe2 silicon photonic modulator with ultra-high stability. *Photon. Res.* **8**, 1189-1196 (2020).

19. Wu K, Wang YF, Qiu CY, Chen JP. Thermo-optic all-optical devices based on two-dimensional materials. *Photon. Res.* **6**, C22-C28 (2018).

20. Jianji DONG ZS. 2D materials as a new platform for photonic applications. *Front. Optoelectron.* **13**, 89-90 (2020).

21. Dong P, *et al.* Thermally tunable silicon racetrack resonators with ultralow tuning power. *Opt. Express* **18**, 20298-20304 (2010).

22. Turning loss into gain. *Nat. Photon.* **11**, 741-741 (2017).

23. Peng B, *et al.* Loss-induced suppression and revival of lasing. *Science* **346**, 328-332 (2014).

24. Guo A, *et al.* Observation of PT-symmetry breaking in complex optical potentials. *Phys. Rev. Lett* **103**, 093902 (2009).

25. Ozdemir SK, Rotter S, Nori F, Yang L. Parity-time symmetry and exceptional points in photonics. *Nat. Mater.* **18**, 783-798 (2019).

26. El-Ganainy R, Makris KG, Khajavikhan M, Musslimani ZH, Rotter S, Christodoulides DN. Non-Hermitian physics and PT symmetry. *Nat.Phys.* **14**, 11 (2018).

27. Zhang J, Li L, Wang G, Feng X, Guan B-O, Yao J. Parity-time symmetry in wavelength space within a single spatial resonator. *Nat. Commun.* **11**, 3217 (2020).

28. Zhang J, Yao J. Parity-time-symmetric optoelectronic oscillator. *Sci Adv* **4**, eaar6782 (2018).

29. Peng B, *et al.* Chiral modes and directional lasing at exceptional points. *Proceedings of the National Academy of Sciences of the United States of America* **113**, 6845-6850 (2016).

30. Brandstetter M, *et al.* Reversing the pump dependence of a laser at an exceptional point. *Nat. Commun.* **5**, 4034 (2014).

31. Hodaei H, Miri MA, Heinrich M, Christodoulides DN, Khajavikhan M. Parity-time-symmetric microring lasers. *Science* **346**, 975-978 (2014).

32. Fleury R, Sounas D, Alu A. An invisible acoustic sensor based on parity-time symmetry. *Nat. Commun.* **6**, 5905 (2015).

33. Lin Z, Ramezani H, Eichelkraut T, Kottos T, Cao H, Christodoulides DN. Unidirectional invisibility induced by PT-symmetric periodic structures. *Phys. Rev. Lett* **106**, 213901 (2011).

34. Regensburger A, Bersch C, Miri MA, Onishchukov G, Christodoulides DN, Peschel U. Parity-time synthetic photonic lattices. *Nature* **488**, 167-171 (2012).

35. Quiroz-Juárez MA, *et al.* Exceptional points of any order in a single, lossy waveguide beam splitter by photon-number-resolved detection. *Photon. Res.* **7**, 862 (2019).

36. Yoon JW, *et al.* Time-asymmetric loop around an exceptional point over the full optical communications band. *Nature* **562**, 86-90 (2018).

37. Xu H, Mason D, Jiang L, Harris JG. Topological energy transfer in an optomechanical system with exceptional points. *Nature* **537**, 80-83 (2016).

38. Doppler J, *et al.* Dynamically encircling an exceptional point for asymmetric mode switching. *Nature* **537**, 76-79 (2016).

39. Liu Q, *et al.* Efficient Mode Transfer on a Compact Silicon Chip by Encircling Moving Exceptional Points. *Phys. Rev. Lett.* **124**, 153903 (2020).

40. Liu Y, *et al.* Highly Efficient Silicon Photonic Microheater Based on Black Arsenic–Phosphorus. *Adv. Opt. Mater.* **8**, 1901526 (2020).

41. Yu L, Dai D, He S. Graphene-based transparent flexible heat conductor for thermally tuning nanophotonic integrated devices. *Appl. Phys. Lett.* **105**, 1679 (2014).